\documentclass[
aps, % 
prl,
preprintnumbers,
twocolumn,
superscriptaddress,
nofootinbib,
floatfix,
10pt
]{revtex4-1}
\usepackage{amsfonts,amssymb,stmaryrd,latexsym,amsmath,braket}
\usepackage{graphicx,subfigure}
\usepackage{times}
\usepackage{slashed}
\usepackage{braket}
\usepackage{verbatim}
\usepackage[utf8]{inputenc}

\usepackage{color}
\usepackage[dvipsnames]{xcolor}
\usepackage[backref=false]{hyperref}
\hypersetup{pdftitle={mcg},
 pdfauthor={TYWU},
 unicode=true, bookmarks=true,bookmarksnumbered=false,bookmarksopen=false, breaklinks=false,pdfborder={0 0 1},backref=false,colorlinks=true, allcolors=BlueViolet}
\usepackage{bm}
\usepackage{soul}

\begin{document}
\title{Reaction Cross Sections and {\boldmath$\alpha$}-Cluster Geometry in {\boldmath$^{12}$}C and Be Isotopes}

\author{Tianyu Wu}
\affiliation{School of Physics, Beihang University, Beijing 102206, China}
\author{Baohua Sun}\thanks{bhsun@buaa.edu.cn}
\affiliation{School of Physics, Beihang University, Beijing 102206, China}
\author{Ulf-G. Meißner}
\affiliation{Helmholtz-Institut für Strahlen- und Kernphysik and Bethe Center for Theoretical Physics, Universität Bonn, D-53115 Bonn, Germany}
\affiliation{Institute for Advanced Simulation (IAS-4), Forschungszentrum Jülich, D-52425 Jülich, Germany}
\affiliation{Peng Huanwu Collaborative Center for Research and Education, International Institute for Interdisciplinary and Frontiers, Beihang University, Beijing 100191, China}
\author{Shihang Shen}\thanks{sshen@buaa.edu.cn}
\affiliation{Peng Huanwu Collaborative Center for Research and Education, International Institute for Interdisciplinary and Frontiers, Beihang University, Beijing 100191, China}
\affiliation{School of Physics, Beihang University, Beijing 102206, China}

\begin{abstract}
Reaction cross sections $\sigma_{\rm R}$ are widely used to infer matter radii, yet their sensitivity to nuclear structure beyond radial one-body distributions is less well understood. We combine complete $A$-body nucleon configurations sampled from \textit{ab initio} nuclear lattice effective field theory (NLEFT) with event-by-event Monte Carlo Glauber calculations, thereby retaining the many-body correlations encoded in NLEFT. Using a fixed binary-collision prescription determined by the measured energy- and isospin-dependent total nucleon-nucleon cross sections, the calculations capture the overall magnitudes and energy dependence simultaneously for the available data on $^{12}$C and $^{9}$Be projectiles on carbon and hydrogen. Controlled randomization of angular correlations at fixed matter root-mean-square radius and spherically averaged one-body radial density produces only a weak change in $\sigma_{\rm R}$ for $^{12}$C but approximately a $10\%$ increase for $^{9}\mathrm{Be}+{}^{1}\mathrm{H}$. The calculations also capture the measured rise--plateau--sharp-rise--reduction trend across $^{7,9\text{--}12}$Be, a distinctive pattern reflecting the evolution of cluster and halo structures along the isotopic chain. 
These results show that $\sigma_{\rm R}$ retains sensitivity to intrinsic many-body geometry beyond a single inferred matter radius, opening a route to studies of exotic $\alpha$-cluster geometries and spatial nucleon correlations through reaction cross sections.
\end{abstract}
\maketitle

\paragraph{}{\itshape Introduction.}
Establishing how the spatial structure of atomic nuclei is encoded in reaction observables is a central problem in nuclear physics, particularly for unstable systems beyond the reach of conventional electromagnetic probes. At intermediate and high energies, Glauber theory has been widely used for this purpose. Its application to radioactive-ion-beam measurements led to the identification of halo nuclei~\cite{tanihataPRL1985}, and subsequent Glauber analyses of the reaction cross sections ($\sigma_{\rm R}$) have been extensively used to extract the spatial extent of unstable nuclei~\cite{OzawaNPA2001693}, to investigate nuclear deformation~\cite{TAKECHIPLB2012}, and to study shell evolution~\cite{TanakaPRL2020, OzawaPRL2000, BagchiPRL2020}. Because reaction observables depend directly on nuclear structure input, they also provide sensitive constraints on nuclear structure descriptions and on approaches that seek a unified treatment of structure and reaction~\cite{MinomoPRL2012, HoriuchiPRC2022, ANPLB2024, An2025JPG, PAN2026140487}. 

Recent advances have enabled full Glauber calculations with variational Monte Carlo wave functions, in which the Glauber phase-shift function is evaluated to all orders in nucleon--nucleon ($NN$) multiple scattering~\cite{horiuchiPRC2026,horiuchiPRL2026}. These calculations provide an important benchmark for the Glauber framework, but two sources of uncertainty remain. First, quantitative predictions rely on empirical $NN$ profile functions constrained by the total and elastic $NN$ cross sections and by the ratio of the real to imaginary parts of the forward scattering amplitude~\cite{HoriuchiPRC2007,IbrahimPRC2008}.  
This limitation is particularly visible for a hydrogen target, for which calculations tend to underestimate the measured reaction cross sections at intermediate energies~\cite{moriguchi_prc_2024,horiuchiPRC2026,NagahisaPRC2018}. Second, reducing the nuclear-structure input to one-body densities discards the configuration-level spatial organization of a finite-$A$ system. The optical-limit approximation retains only the first cumulant and therefore depends solely on one-body densities~\cite{Suzuki2003}, whereas a second-order cumulant calculation using microscopic two-body information has recently been reported~\cite{horiuchiPRC2026}. Complete finite-$A$ configurations additionally retain intrinsic cluster geometry and spatial nucleon correlations. The extent to which $\sigma_{\rm R}$ responds to this information at a fixed matter radius and spherically averaged one-body radial density remains unclear.

The Monte Carlo implementation of Glauber theory (MCG) offers a transparent framework for retaining this configuration-level information~\cite{mcg1985,MCG1989,Miller2007}. It samples explicit projectile and target configurations and evaluates binary $NN$ encounters event by event in the transverse plane, allowing local geometry, fluctuations, clustering, and spatial nucleon correlations encoded in the many-body input to enter the reaction calculation directly. The explicit coordinate representation also permits controlled angular rearrangements that preserve the matter radius and one-body radial density, enabling the response of $\sigma_{\rm R}$ to intrinsic geometry to be examined at fixed radial structure. MCG is well established in relativistic heavy-ion collisions, where event-by-event nuclear geometry is related to collective observables~\cite{haojiexuPRL2024,GiacalonePRL2021,zhangcjPRL2022,STARnature2024,STAR2025RPP}, and recent applications of \emph{ab initio} configurations have demonstrated sensitivity to detailed many-body correlations~\cite{GiacalonePRL2025134,GiacalonePRL2025135,DuguetreviewEPJA2025,LiPeiPRL2026,huarxiv2025}. 
Compared with these relativistic flow observables, intermediate-energy reaction cross sections provide a more direct and widely accessible probe of short-lived nuclei without the intervening quark--gluon-plasma evolution; however, their sensitivity to intrinsic many-body geometry within event-by-event MCG has not been systematically explored.

In this Letter, we combine MCG with \emph{ab initio} nucleon configurations from nuclear lattice effective field theory (NLEFT) and compare the calculated reaction cross sections directly with experiment, without first converting the measured $\sigma_{\rm R}$ into model-dependent matter radii. We first benchmark the fixed MCG prescription against available data and Glauber calculations using identical NLEFT structure inputs for $^{12}$C and $^{9}$Be projectiles on carbon and hydrogen.
We then construct controlled configuration ensembles by rotating the projectile nucleons about their center of mass (CoM) while preserving it, the matter radius, and the one-body radial density, thereby isolating the effect of angular correlations and intrinsic cluster geometry. Finally, we apply the same framework to $^{7,9\text{--}12}$Be projectiles on $^{9}$Be and $^{12}$C targets at $790~\mathrm{MeV/nucleon}$ and compare the isotope dependence directly at the cross-section level.

\paragraph{}{\itshape Formalism.} 
The MCG calculation employs the straight-line collision geometry of the
eikonal approximation~\cite{softwarex2015,Loizidesprc2018,
loizidesPRC2026}. For each event, projectile and target nucleon
configurations are sampled and kept frozen during the collision. The impact
parameter $b$ is the transverse separation between the two nuclear centers.
To sample the transverse collision area uniformly, we draw a uniform random
number $u\in[0,1]$ and set $b=b_{\max}\sqrt{u}$. Here, $b_{\max}$ is chosen to be larger than the sum of the projectile and target radii.
At the nucleon--nucleon level, we adopt the hard-sphere prescription: a projectile nucleon $i$ and a target nucleon $j$ collide if $d_{ij}^{\perp}<D$, where $d_{ij}^{\perp}$ is their transverse separation. For a given incident energy and isospin channel, the collision distance $D=\sqrt{\sigma_{NN}^{\mathrm{tot}}/\pi}$ is fixed by the corresponding total $NN$ cross section $\sigma_{NN}^{\mathrm{tot}}$.
Based on the sensitivity test presented in the Supplemental Material~\cite{SM}, we adopt the hard-sphere prescription as the appropriate choice for the present study.
The corresponding $pp$ and $pn$ total cross sections are taken from the PDG compilation~\cite{PDG2024} and are also used in the previous Glauber calculation~\cite{ZHANGSB2024,jianweiPLB2024}, so that both approaches employ identical $NN$ cross-section inputs. At a given incident energy, these cross sections completely specify
the binary-collision criterion, introducing no additional adjustable reaction parameters. An event containing at least one binary $NN$ collision is classified as a reaction event, and $\sigma_{\rm R}$ is obtained
by averaging over the sampled configurations and transverse collision area~\cite{SM}.
MCG therefore implements the reaction directly at the probability level
through binary $NN$ collisions~\cite{Miller2007}, without constructing the
complex many-body Glauber $S$ matrix used in coherent-amplitude
formulations~\cite{IbrahimPRC2008}.

In this work, we employ \emph{ab initio} configurations from nuclear
lattice effective field theory (NLEFT)~\cite{Elhatisari:2017eno,
shenNC2023, ShenPRL2025}. In NLEFT, effective field theory is formulated on a discretized space--time lattice and solved
with Monte Carlo methods to compute properties of nuclear many-body
systems~\cite{Lee:2025req, Lahde:2019npb}. Using wave-function matching to alleviate the sign problem, calculations with high-fidelity
next-to-next-to-next-to-leading-order (N$^3$LO) chiral interactions are
organized around a sign-friendly Hamiltonian, with the remaining interaction terms included perturbatively, and provide a successful description of binding energies and radii from light to medium-heavy
nuclei~\cite{Serdarnature2024}.
For the Be isotopes considered here, the N$^3$LO ground-state
energies agree with experiment typically around $\sim$1~MeV and the matter radii
capture the halo enhancement of $^{11}$Be, $r_m=2.86(1)$~fm versus $2.91(5)$~fm experimentally~\cite{ShenPRL2025}.
The full $A$-body coordinate correlations enter the reaction calculation through the pinhole algorithm~\cite{Elhatisari:2017eno}, which samples
nucleon coordinates according to the amplitude of the many-body density operator,
\begin{equation}
Z=\langle\Psi|M^{L_t/2}\rho(\mathbf{n}_1,\ldots,\mathbf{n}_A)
M^{L_t/2}|\Psi\rangle,
\end{equation}
where $M$ is the transfer matrix, $\Psi$ the trial wave function, $\rho(\mathbf{n}_1,\ldots,\mathbf{n}_A)$ the normal-ordered product of single-nucleon density operators, and observables depending on $\{\mathbf{n}\}$ are obtained by averaging over the same ensemble.
We integrate these NLEFT configurations into the MCG framework and propagate the associated perturbative corrections to the reaction observable itself. Specifically, the event-level MCG reaction indicator is evaluated as a many-body observable within the perturbative NLEFT estimator, so that the corrected finite-$A$ configuration weights of both projectile and target nuclei enter the event average directly. This retains the many-body correlations encoded in the NLEFT ensemble, rather than reducing the structure input to one-body densities. Details of the formulation are given in Ref.~\cite{SM}.

\paragraph{\itshape Results and discussion.} 
\begin{figure}[h] 
\centering 
\includegraphics[width=0.5\textwidth]{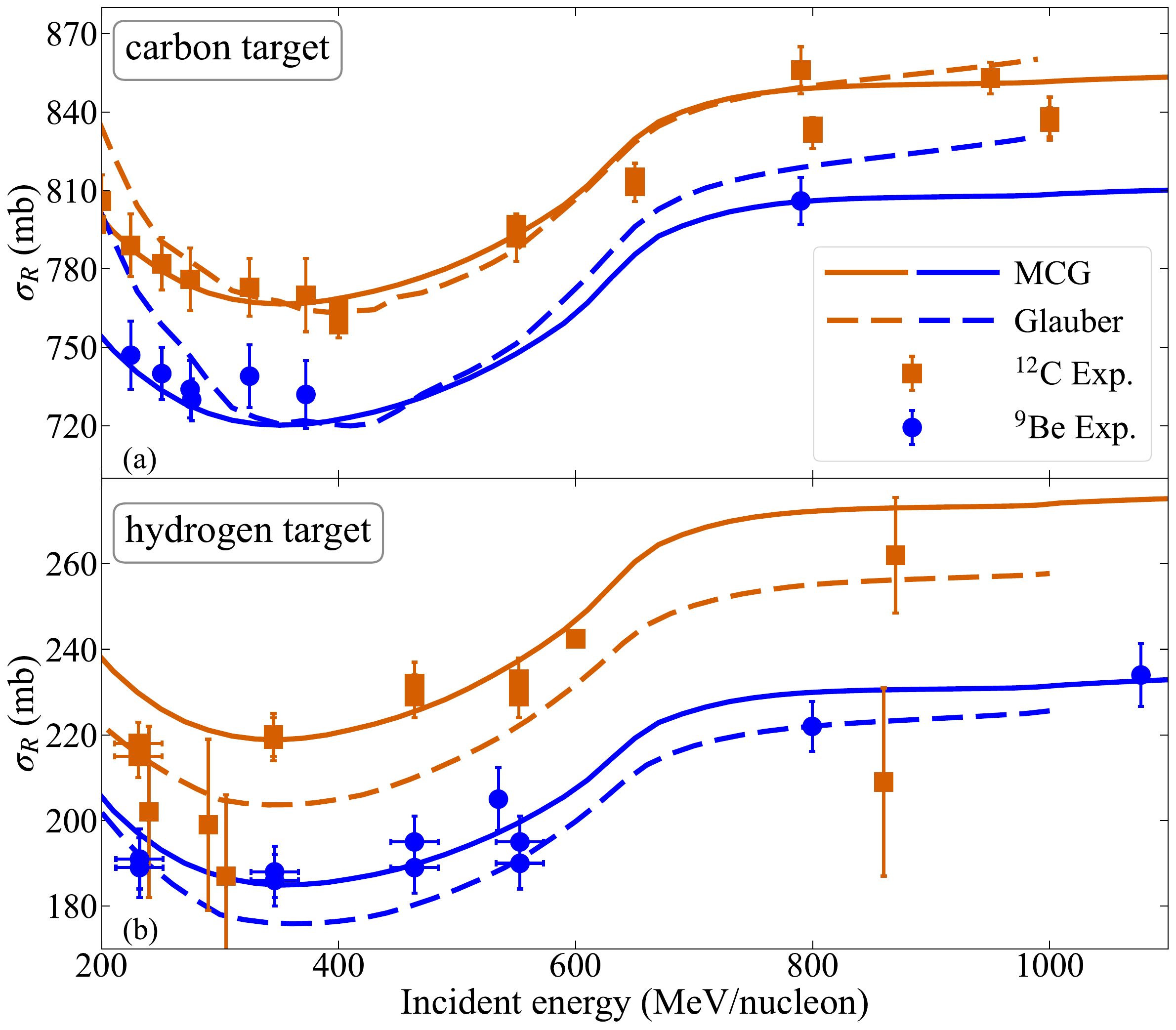} 
\caption{Reaction cross sections for $^{12}$C (orange squares) and $^{9}$Be (blue circles) projectiles on carbon (a) and hydrogen (b) targets. Symbols denote experimental data~\cite{12con12CKOX1987PRC,OzawaNPA2001693,12Con12CPONNATH2024PLB,12con12CTakechi2009PRC,12ConPJaros1978PRC,12ConPRENBERG1972NPA,12ConPWebber1990PRC,CarlsonADNDT1996,Dietrich2002}; solid and dashed curves show the corresponding MCG and Glauber calculations, respectively. Monte Carlo statistical uncertainties are smaller than the line widths.
}\label{fig:1} 
\end{figure}  

The theoretical and experimental reaction cross sections for $^{12}$C and
$^{9}$Be projectiles on carbon and hydrogen targets are compared in
Fig.~\ref{fig:1}. The curves labeled ``Glauber'' are obtained from Glauber calculations, in which the phase-shift function is evaluated with all orders of $NN$ multiple scattering retained, following the treatment of Refs.~\cite{horiuchiPRC2026,horiuchiPRL2026}. For each reaction system, the Glauber and MCG calculations use identical nuclear structure inputs, namely, the same NLEFT configuration ensembles for both projectile and target. Fig.~\ref{fig:1} therefore provides a controlled comparison between the two reaction treatments with the nuclear structure input held fixed. The Glauber curves are shown up to $1000$~MeV/nucleon, the upper limit of the profile-function parameter set tabulated in Ref.~\cite{IbrahimPRC2008}.

For the carbon target, Fig.~\ref{fig:1}(a) shows that the MCG and Glauber calculations give comparable overall magnitudes for both $^{12}$C and $^{9}$Be projectiles and are broadly consistent with the measurements. Their differences become visible mainly around $200$--$300$~MeV/nucleon and above $800$~MeV/nucleon. In particular, from $800$ to $1000$~MeV/nucleon, the empirical $\sigma_{NN}^{\mathrm{tot}}$ input shows no corresponding upward trend. The MCG results remain nearly constant, consistent with the available data, whereas the Glauber results increase. Since identical nuclear-structure and $\sigma_{NN}^{\mathrm{tot}}$ inputs are employed, this increase cannot be attributed to $\sigma_{NN}^{\mathrm{tot}}$ alone and instead illustrates the sensitivity of the calculation to the additional range and phase parameters entering the complex profile function~\cite{IbrahimPRC2008}.

For the hydrogen target, the separation between the two treatments becomes
more pronounced, as shown in Fig.~\ref{fig:1}(b). For both projectiles, MCG systematically yields larger reaction cross sections than the Glauber calculation. Glauber calculations have previously been reported to lie below the measured cross sections for proton--nucleus systems
at intermediate energies~\cite{horiuchiPRC2026,moriguchi_prc_2024}. Thus, the
systematic increase from Glauber to MCG is in the direction
suggested by those comparisons. Given the sizable
experimental uncertainties, we assess the agreement using the chi-square per
datum, defined as $\chi^2/N=N^{-1}\sum_{i=1}^{N}[(\sigma_{R,i}^{\mathrm{th}}-\sigma_{i}^{\mathrm{exp}})/
\Delta\sigma_{i}^{\mathrm{exp}}]^2$. For the experimental data in Fig.~\ref{fig:1} (b) at $E\geq300~\mathrm{MeV/nucleon}$, $\chi^2/N$ is $1.5$ for MCG and $5.1$ for the Glauber calculation. According to this uncertainty-weighted measure, MCG gives a better overall agreement with the hydrogen target data.

Taken together, MCG more closely reproduces the weak high-energy dependence observed for the carbon target and yields a substantially smaller $\chi^2/N$ for the hydrogen target than the Glauber calculation.
These comparisons show that the same fixed MCG prescription provides a reasonable description of the systems and energies considered here without system-dependent retuning. MCG therefore provides a direct and less parametrization-dependent connection between elementary $NN$ scattering and nuclear reaction cross sections.

\begin{figure*}[!t] 
\centering 
\includegraphics[width=1\textwidth]{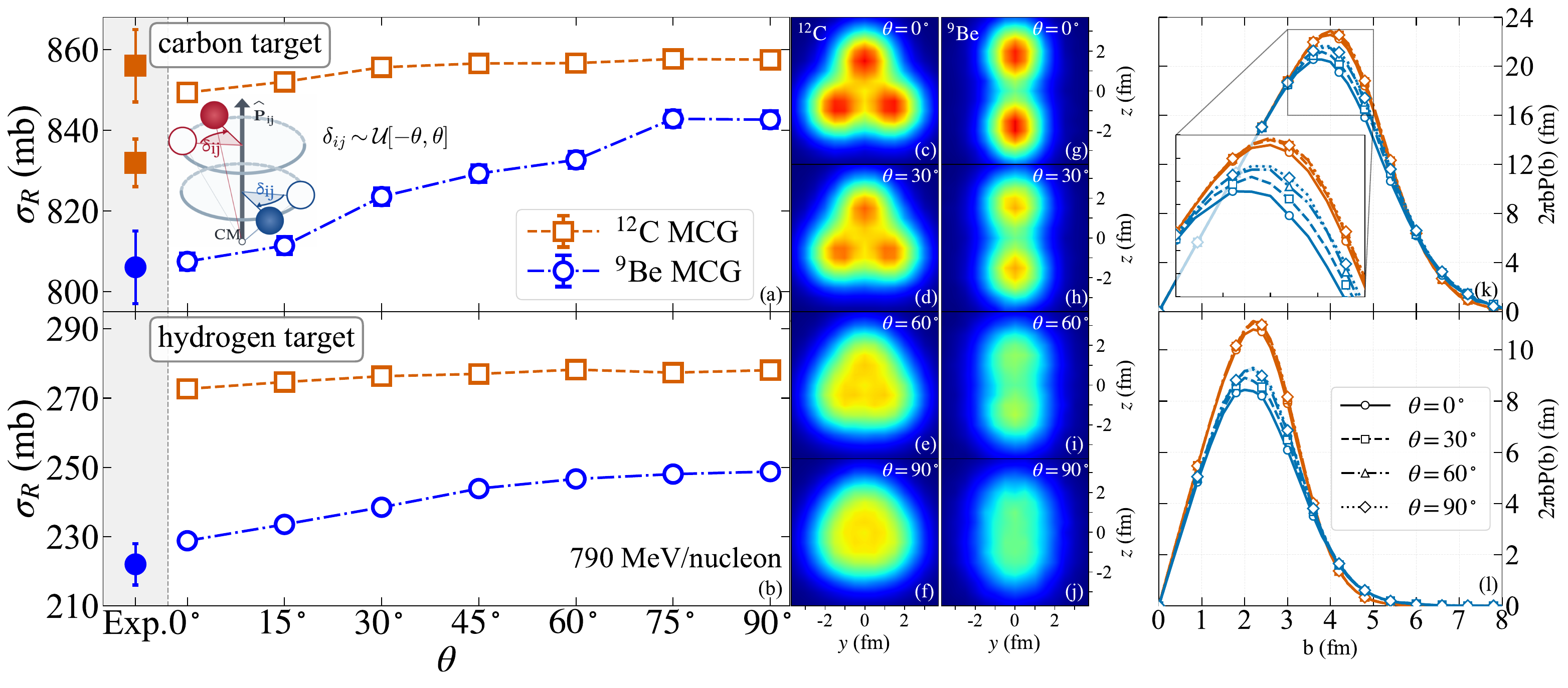} 
\caption{Controlled angular decorrelation of projectile configurations at $790~\mathrm{MeV/nucleon}$. The schematic inset in panel (a) shows a randomly selected nucleon pair before (open spheres) and after (solid spheres) a common rotation about $\mathbf P_{ij}$. Here $\delta_{ij}\sim\mathcal U[-\theta,\theta]$ is the signed angle from $\mathbf r_{k\perp}$ to $\mathbf r_{k\perp}'$ ($k=i,j$). Panels (a) and (b) show the MCG reaction cross sections $\sigma_{\rm R}$ for $^{12}$C and $^{9}$Be projectiles on carbon and hydrogen targets, respectively, as functions of $\theta$. Panels (c)--(f) and (g)--(j) display the aligned intrinsic-density slices in the $yz$ plane for $^{12}$C and $^{9}$Be, respectively, at $\theta=0^\circ$, $30^\circ$, $60^\circ$, and $90^\circ$. Panels (k) and (l) show the corresponding reaction probabilities $2\pi \mathrm{bP(b)}$ for the two targets; the inset in panel (k) enlarges the boxed region.}
\label{fig:alpha_decorrelation}
\end{figure*}

To isolate angular many-body correlations while keeping the radial one-body structure fixed, we generate auxiliary projectile ensembles through constrained rotations of randomly selected nucleon pairs. 
Here, $\mathbf r_k$ denotes the position of projectile nucleon $k$ relative to the projectile CoM, taken as the origin. For each pair $(i,j)$, we define
$\mathbf P_{ij}=\mathbf r_i+\mathbf r_j$ and rotate both position
vectors by the same signed angle
$\delta_{ij}\sim\mathcal{U}[-\theta,\theta]$ about the axis along
$\mathbf P_{ij}$ and passing through the projectile CoM:
\begin{equation}
  \mathbf r_k'
  =
  \mathcal R_{\widehat{\mathbf P}_{ij}}(\delta_{ij})\mathbf r_k,
  \qquad k=i,j,
  \label{eq:pair_rotation}
\end{equation}
where $\widehat{\mathbf P}_{ij}=\mathbf P_{ij}/|\mathbf P_{ij}|$.
Because $\mathbf P_{ij}$ lies along the rotation axis,
$\mathbf r_i'+\mathbf r_j'
=\mathcal R_{\widehat{\mathbf P}_{ij}}(\delta_{ij})
(\mathbf r_i+\mathbf r_j)=\mathbf P_{ij}$, so the pair contribution to the projectile CoM remains unchanged. 
Applying this construction to every pair in the random partition (with one nucleon left unchanged for odd $A$) preserves the total coordinate sum and hence the projectile CoM.
The rotation also preserves $|\mathbf r_k'|=|\mathbf r_k|$, leaving
every single-particle radius, the matter radius, and the spherically
averaged one-body radial density unchanged, while rearranging the angular configuration of the pair. Repeating the procedure with new random pairs progressively
decorrelates the nucleon directions within each configuration, thereby
suppressing its original many-body angular correlations.
Only the projectile is modified, while the target retains its original NLEFT
correlations. The parameter $\theta$ controls the sampled range of the
pair-rotation phase, with $\theta=0^\circ$ recovering the original NLEFT
ensemble. The configurations in
Figs.~\ref{fig:alpha_decorrelation}(c)--(j) are aligned only to visualize
the resulting intrinsic structures; the reaction calculations average
uniformly over the global projectile orientation. 

The $^{12}$C ground state is described in NLEFT by a compact, approximately triangular three-$\alpha$ configuration~\cite{EpelbaumPRL2012,shenNC2023}. This structure appears as three localized $\alpha$-rich peaks in the unmodified NLEFT ensemble, as shown in Fig.~\ref{fig:alpha_decorrelation}(c). As $\theta$ increases, the peaks progressively broaden and merge in Figs.~\ref{fig:alpha_decorrelation}(c)--(f), whereas both $2\pi \mathrm{bP(b)}$ and $\sigma_{\rm R}$ change only weakly in Figs.~\ref{fig:alpha_decorrelation}(a), (b), (k), and (l). In the triangular configuration, the three $\alpha$-rich regions extend along different directions from the CoM and are therefore relatively dispersed within the intrinsic plane. For both targets, $\sigma_{\rm R}$ remains unchanged within statistical uncertainties from $\theta=75^\circ$ to $90^\circ$. Additional calculations at $\theta>90^\circ$ (not shown) agree with the $\theta=90^\circ$ results, showing that the response of $\sigma_{\rm R}$ to further angular decorrelation has reached a plateau.

The $^{9}$Be ground state instead exhibits a molecular
$2\alpha+n$ structure, in which two $\alpha$-rich regions form a
pronounced dumbbell-like distribution along a single dominant
axis~\cite{ShenPRL2025}, as shown in
Fig.~\ref{fig:alpha_decorrelation}(g). As $\theta$ increases, the two
peaks broaden, the intervening neck is progressively filled, and the
aligned density becomes shorter along the original
$\alpha$--$\alpha$ axis and wider in the perpendicular direction in
Figs.~\ref{fig:alpha_decorrelation}(g)--(j). 
This angular redistribution enhances $2\pi \mathrm{bP(b)}$, the reaction probability at impact parameter $b$, mainly at $b\simeq3$--$5~\mathrm{fm}$ for the carbon target and at $b\simeq1.5$--$3~\mathrm{fm}$ for the hydrogen target, as shown in Figs.~\ref{fig:alpha_decorrelation}(k) and (l). 
At smaller $b$, strong overlap suppresses the response to angular decorrelation, particularly for the carbon target as $\mathrm{P(b)}$ approaches saturation~\cite{MakiguchiPTEP2022}. At large $b$, the reaction probes the unchanged radial tail, causing the curves to merge~\cite{IbrahimPRC2008}. The influence of angular correlations is therefore concentrated in the intervening grazing region.
When integrated over $b$, these enhancements increase $\sigma_{\rm R}$ from $\theta=0^\circ$ to $90^\circ$ by approximately $35$ mb ($4\%$) for $^{9}$Be on carbon and $20$ mb ($9\%$) for $^{9}$Be on hydrogen, as shown in Figs.~\ref{fig:alpha_decorrelation}(a) and (b).

The contrasting responses reveal a geometry-selective sensitivity of $\sigma_{\rm R}$ to the angular organization of $\alpha$ clustering. Within each $\theta$ scan, the matter radius and spherically averaged one-body radial density are unchanged, so the $\theta$-dependent variation of $\sigma_{\rm R}$ reflects changes in angular organization rather than in radial size or the radial tail. The compact triangular three-$\alpha$ structure of $^{12}$C distributes its $\alpha$-rich regions along three directions in the intrinsic plane, whereas the dumbbell-like $2\alpha+n$ structure of $^{9}$Be concentrates its two $\alpha$-rich regions near opposite ends of a single dominant axis. This controlled angular decorrelation therefore produces a much weaker response in $^{12}$C than in $^{9}$Be, showing that $\sigma_{\rm R}$ distinguishes how the $\alpha$-rich regions are organized even when the radial one-body structure is held fixed within each comparison.

\begin{figure}[h] 
\centering 
\includegraphics[width=0.4\textwidth]{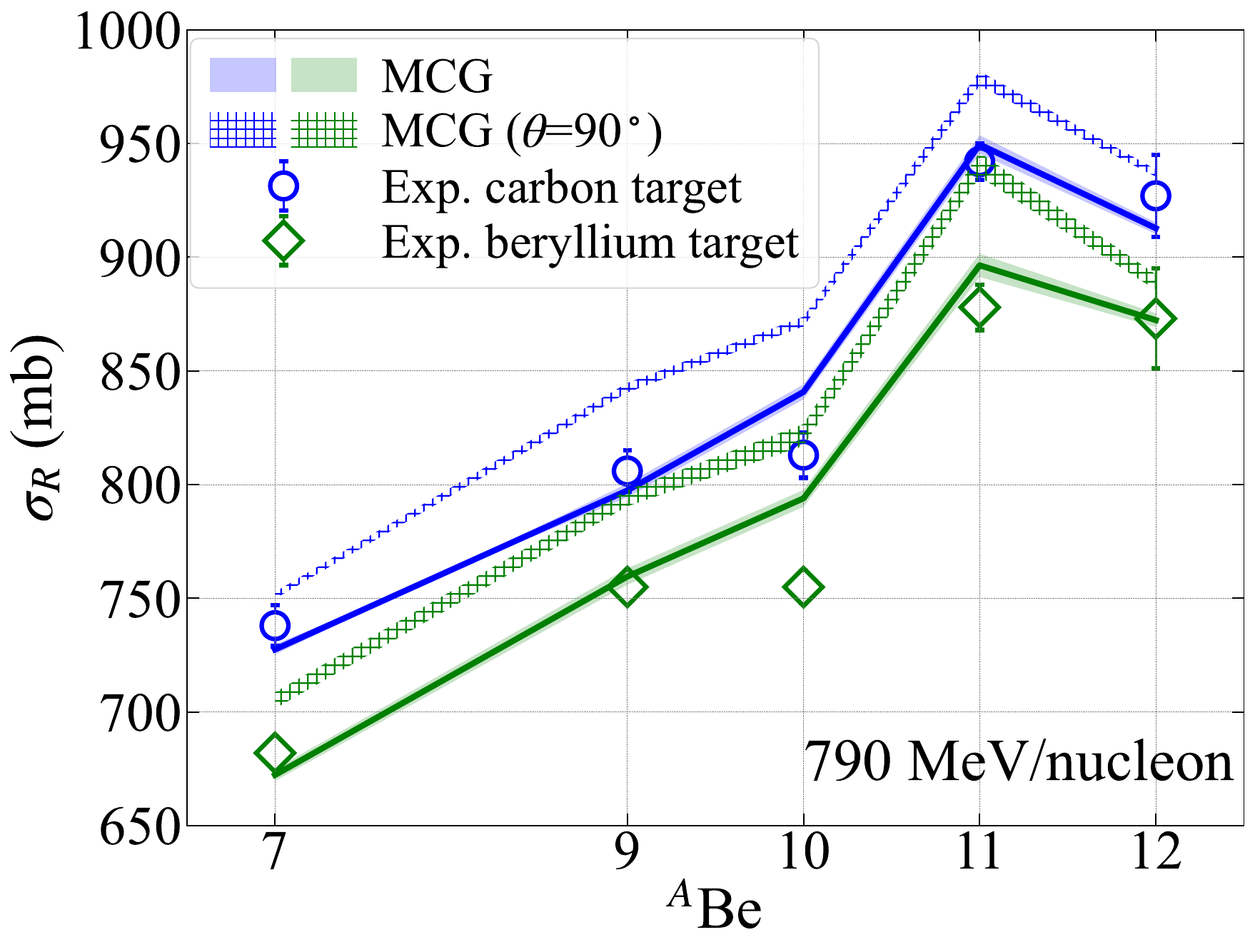} 
\caption{Reaction cross sections $\sigma_{\rm R}$ of
$^{7,9\text{--}12}$Be projectiles on beryllium and carbon
targets at $790~\mathrm{MeV/nucleon}$. 
The correlated NLEFT$+$MCG calculations and the results obtained after applying the pair-rotation procedure with $\theta=90^\circ$ are compared
with the experimental data~\cite{tanihataPRL1985,
Beon9Be12CTANIHATA1988PLB}.
}\label{fig:Be_RCS}
\end{figure}

We next apply the same controlled angular decorrelation procedure to the
$^{7,9\text{--}12}$Be isotopic chain, which spans clustered,
molecular-orbital, and halo structures. Fig.~\ref{fig:Be_RCS} compares
the MCG results obtained from the original correlated
configurations with those obtained after angular decorrelation at
$\theta=90^\circ$. For every isotope and on both targets, angular
decorrelation increases $\sigma_{\rm R}$ while leaving the overall stepwise
isotope trend qualitatively intact. 
The enhancements amount to approximately $24$--$45$ mb ($3\%$--$6\%$) for the carbon target and $17$--$44$ mb ($2\%$--$5\%$) for the beryllium target.
These changes exceed the approximately $1\%$ precision currently achievable in experiments.
Since the rotations preserve the matter radius and spherically averaged one-body radial density, these shifts demonstrate a percent-level response of $\sigma_{\rm R}$ to many-body angular
correlations at fixed radial structure, with the magnitude depending on
both the isotope and the target. A deformed Glauber analysis using densities from NLEFT inputs also found a smaller $\sigma_{\rm R}$ when orientation dependence was included than for spherical averages~\cite{lu2026deformation}.

Across the Be isotopes considered here, the calculated cross sections reflect the structural differences encoded in the NLEFT configurations used as reaction inputs~\cite{ShenPRL2025}.
From $^{7}$Be to $^{9}$Be, the intrinsic density evolves from a configuration resembling $^{4}$He+$^{3}$He with one
prominent $\alpha$-rich region to a pronounced two-center $2\alpha+n$
structure, while the matter radius increases from $2.39$ to $2.52$~fm.
These changes in intrinsic geometry and spatial extent are reflected in
the first marked rise in $\sigma_{\rm R}$. From $^{9}$Be to $^{10}$Be, the
two-$\alpha$ core is largely preserved, the valence neutrons
predominantly occupy $\pi$-type orbitals, and the matter radius changes little; correspondingly, $\sigma_{\rm R}$ forms a near plateau. NLEFT
underbinds $^{10}$Be by about $1.3$~MeV and overestimates its matter
radius by approximately $10\%$. 
The overly extended NLEFT density is inherited by the reaction calculation and likely contributes to the remaining overestimate of $\sigma_{\rm R}$. The discrepancies in the energy and matter radius point to the need for further improvements in the three-nucleon interaction, as also emphasized in a recent $n$--$\alpha$ scattering study~\cite{Elhatisari:2025fyu}.
From $^{10}$Be to $^{11}$Be, NLEFT develops
a spatially extended $\sigma$-type valence-neutron distribution with
strong $s$-wave character~\cite{shenparticles2026}, increasing the matter radius to $2.86$~fm;
this structural change is reflected in the sharp rise in $\sigma_{\rm R}$.
From $^{11}$Be to $^{12}$Be, the density becomes more rounded and
compact and the matter radius decreases to $2.63$~fm, leading to a
modest reduction in $\sigma_{\rm R}$.

The correspondence between the calculated and measured rise--plateau--sharp-rise--reduction patterns shows how the structural evolution encoded in the NLEFT inputs is reflected in the reaction cross sections. Rather than first extracting matter rms radii from the
measured $\sigma_{\rm R}$---a model-dependent mapping of the reaction observable onto a single size parameter that may obscure information beyond overall nuclear size---we compare the measured and calculated isotope trends directly at the cross-section level. The structural picture manifested in this cross-section evolution is consistent with
independent spectroscopic evidence~\cite{YangPRL2014,ChenPRL2025} and
microscopic structure calculations~\cite{ZhaoPRC2022,CaiPRC2025,
OdsurenPRC2026}. This correspondence indicates that reaction cross
sections retain sensitivity to subtle many-body structural evolution
and contain information beyond that captured by a single rms radius.

\paragraph{Summary.}
We have combined complete $A$-body nucleon configurations sampled from \textit{ab initio} NLEFT many-body wave functions with event-by-event MCG calculations, allowing the many-body correlations encoded in the wave functions to enter $\sigma_{\rm R}$ directly rather than reducing the nuclear structure input to densities. Using the same fixed binary-collision prescription for all systems, the calculations capture the overall magnitudes and energy dependence of the available data for $^{12}$C and $^{9}$Be projectiles on carbon and hydrogen without system-dependent retuning. Controlled angular decorrelations that preserve the matter radius and one-body radial density change $\sigma_{\rm R}$ only weakly for the compact, approximately triangular three-$\alpha$ structure of $^{12}$C, but increase it by approximately $4\%$ for $^{9}$Be on carbon and $9\%$ for $^{9}$Be on hydrogen, reflecting the pronounced dumbbell-like $2\alpha+n$ geometry of $^{9}$Be. Applied across $^{7,9\text{--}12}$Be, the same framework also captures the measured rise--plateau--sharp-rise--reduction trend, reflecting the evolution of cluster and halo structures encoded in the NLEFT configurations. These results show that $\sigma_{\rm R}$ is not determined by a single matter radius or radial one-body density alone, but retains sensitivity to intrinsic $\alpha$ cluster many-body geometry. Direct cross-section comparisons can preserve structural information obscured when $\sigma_{\rm R}$ is reduced to a single inferred radius. More broadly, the same framework can be extended to study other short-lived nuclei to test exotic $\alpha$-cluster geometries and spatial nucleon correlations.

\begin{acknowledgements}
\paragraph{Acknowledgements}
We are grateful for insightful discussions with Serdar Elhatisari, Ye-Lei Sun, Ya-Kun Wang, and Lin-Qian Wu. This work was partly supported by the National Natural Science Foundation of China (Nos. 12325506, U2541242, and 12435007) and by ``the Fundamental Research Funds for the Central Universities''. We gratefully acknowledge the Computational resources provided by the HPC platform of Beihang University. The work of UGM was supported in part by the European
Research Council (ERC) under the European Union's Horizon 2020 research
and innovation program (EXOTIC, grant agreement No. 101018170),
and by the CAS President's International Fellowship Initiative (PIFI) (Grant No.~2025PD0022). 
\end{acknowledgements}

\bibliography{References}
\bibliographystyle{apsrev} 
\end{document}

\section*{SUPPLEMENTAL MATERIAL}
\addtocounter{section}{10}

\subsection{Monte Carlo Glauber model}
\label{sec:sm_mcg}

In a Monte Carlo Glauber (MCG) calculation, each event is
specified by a finite-$A$ coordinate configuration of the projectile
and target. We denote the sampled nucleon coordinates by
\begin{equation}
  \mathbf R_P^{(e)}
  =
  \{\mathbf r_{P,1}^{(e)},\ldots,\mathbf r_{P,A_P}^{(e)}\},
  \qquad
  \mathbf R_T^{(e)}
  =
  \{\mathbf r_{T,1}^{(e)},\ldots,\mathbf r_{T,A_T}^{(e)}\}.
  \label{eq:sm_mcg_config}
\end{equation}
Before evaluating the collision, the projectile and target
configurations are re-centered to their own total centers of mass and
then embedded in a common collision frame with impact-parameter
vector $\mathbf b$.  The transverse coordinates of the projectile and
target nucleons are denoted by $\mathbf s_{P,i}^{(e)}$ and
$\mathbf s_{T,j}^{(e)}$, respectively.  For a given event and impact
parameter, the transverse separation of projectile nucleon $i$ and
target nucleon $j$ is
\begin{equation}
  b_{ij}^{(e)}(\mathbf b)
  =
  \left|
  \mathbf s_{P,i}^{(e)}
  +
  \mathbf b
  -
  \mathbf s_{T,j}^{(e)}
  \right|.
  \label{eq:sm_pair_separation}
\end{equation}

In the hard-sphere implementation used in the present work, a
projectile nucleon $i$ and a target nucleon $j$ are counted as a
binary collision when
\begin{equation}
  b_{ij}^{(e)}(\mathbf b)
  <
  \sqrt{\frac{\sigma_{ij}^{NN}}{\pi}},
  \label{eq:sm_binary_condition}
\end{equation}
where $\sigma_{ij}^{NN}$ is the energy- and isospin-dependent
elementary nucleon--nucleon total cross section used in the reaction
calculation.  Equivalently, we introduce the pair-collision indicator
\begin{equation}
  C_{ij}^{(e)}(\mathbf b)
  =
  \Theta
  \left[
  \frac{\sigma_{ij}^{NN}}{\pi}
  -
  \left(
  b_{ij}^{(e)}(\mathbf b)
  \right)^2
  \right].
  \label{eq:sm_pair_indicator}
\end{equation}
An event is counted as a reaction event if at least one binary 
nucleon--nucleon collision occurs.  The corresponding event-level
reaction indicator is
\begin{equation}
  I_R^{(e)}(\mathbf b)
  =
  1-
  \prod_{i=1}^{A_P}
  \prod_{j=1}^{A_T}
  \left[
  1-
  C_{ij}^{(e)}(\mathbf b)
  \right].
  \label{eq:sm_mcg_event_indicator}
\end{equation}
The reaction probability $P_R^{\rm MCG}$ at impact-parameter magnitude
$b=|\mathbf b|$ is obtained by averaging this event-level indicator
over the projectile and target configurations, their orientations, and
the azimuthal direction of $\mathbf b$,
\begin{equation}
  P_R^{\rm MCG}(b)
  =
  \left\langle
  I_R^{(e)}(\mathbf b)
  \right\rangle_e .
  \label{eq:sm_mcg_probability_average}
\end{equation}
The reaction cross section is then obtained directly as
\begin{equation}
  \sigma_R^{\rm MCG}
  =
  \int d^2b\,
  P_R^{\rm MCG}(b)
  =
  2\pi
  \int_0^\infty b\,db\,
  P_R^{\rm MCG}(b).
  \label{eq:sm_mcg_sigma}
\end{equation}
Thus, once the event-level reaction indicator in
Eq.~\eqref{eq:sm_mcg_event_indicator} has been constructed, no
complex many-body $S$ matrix is required in the MCG calculation.
The microscopic calculations reported in the main text use the full
NLEFT many-body configurations directly as the structural input.

\subsubsection{Dependence on the nucleon--nucleon collision profile}
\label{sec:sm_nn_profile}

The binary-collision prescription adopted in the present MCG calculation
corresponds to a hard-sphere nucleon--nucleon ($NN$) collision profile.
To examine the sensitivity to this assumption, we also consider the
smooth profile~\cite{RybczynskiJPG2014,Loizidesprc2018}
\begin{equation}
  p_{ij}^{(\omega)}(b_{ij})
  =
  \frac{
    \Gamma\left(
      1/\omega,\,
      \frac{b_{ij}^{2}}{D_{ij}^{2}\omega}
    \right)
  }{
    \Gamma(1/\omega)
  },
  \qquad
  D_{ij}
  =
  \sqrt{\frac{\sigma_{ij}^{NN}}{\pi}},
  \label{eq:sm_omega_profile}
\end{equation}
Here, $\Gamma(a,x)=\int_x^\infty t^{a-1}e^{-t}\,dt$ denotes the
upper incomplete Gamma function, and
$\Gamma(a)\equiv\Gamma(a,0)$ is the complete Gamma function.
Equation~\eqref{eq:sm_omega_profile} has the same functional form
as the collision profiles used in
Refs.~\cite{RybczynskiJPG2014,Loizidesprc2018}, where the transverse
profile is conventionally normalized to the inelastic (wounding)
$NN$ cross section. In the present work, we retain this functional
form but instead set its normalization by the energy- and
isospin-dependent total $NN$ cross section:
\begin{equation}
  \int d^2b_{ij}\,
  p_{ij}^{(\omega)}(b_{ij})
  =
  2\pi\int_0^\infty
  b_{ij}\,db_{ij}\,
  p_{ij}^{(\omega)}(b_{ij})
  =
  \pi D_{ij}^{\,2}
  =
  \sigma_{ij}^{NN} .
  \label{eq:sm_profile_normalization}
\end{equation}
Accordingly, $p_{ij}^{(\omega)}$ is interpreted here as an
effective pair-encounter probability in the composite-nucleus
reaction calculation, rather than as the standard free-$NN$
inelastic wounding probability. The parameter $\omega$ controls
the diffuseness of the transverse collision profile. The
hard-sphere prescription is recovered in the limit
$\omega\rightarrow 0$.
For the smooth profile, the corresponding MCG reaction probability is
\begin{equation}
P_R^{\rm MCG}(b;\omega)
=
\left\langle
1-
\prod_{i=1}^{A_P}
\prod_{j=1}^{A_T}
\left[
1-p_{ij}^{(\omega)}
\left(b_{ij}^{(e)}(\mathbf b)\right)
\right]
\right\rangle_e ,
\qquad b=|\mathbf b|.
\label{eq:sm_omega_event_probability}
\end{equation}
For a fixed projectile--target configuration, the expression inside
$\langle\cdots\rangle_e$ is the conditional probability that at least
one binary collision occurs. This construction assumes independent
binary $NN$ encounters conditional on the sampled geometry.

\begin{figure}[t]
  \centering
  \includegraphics[width=0.8\linewidth]{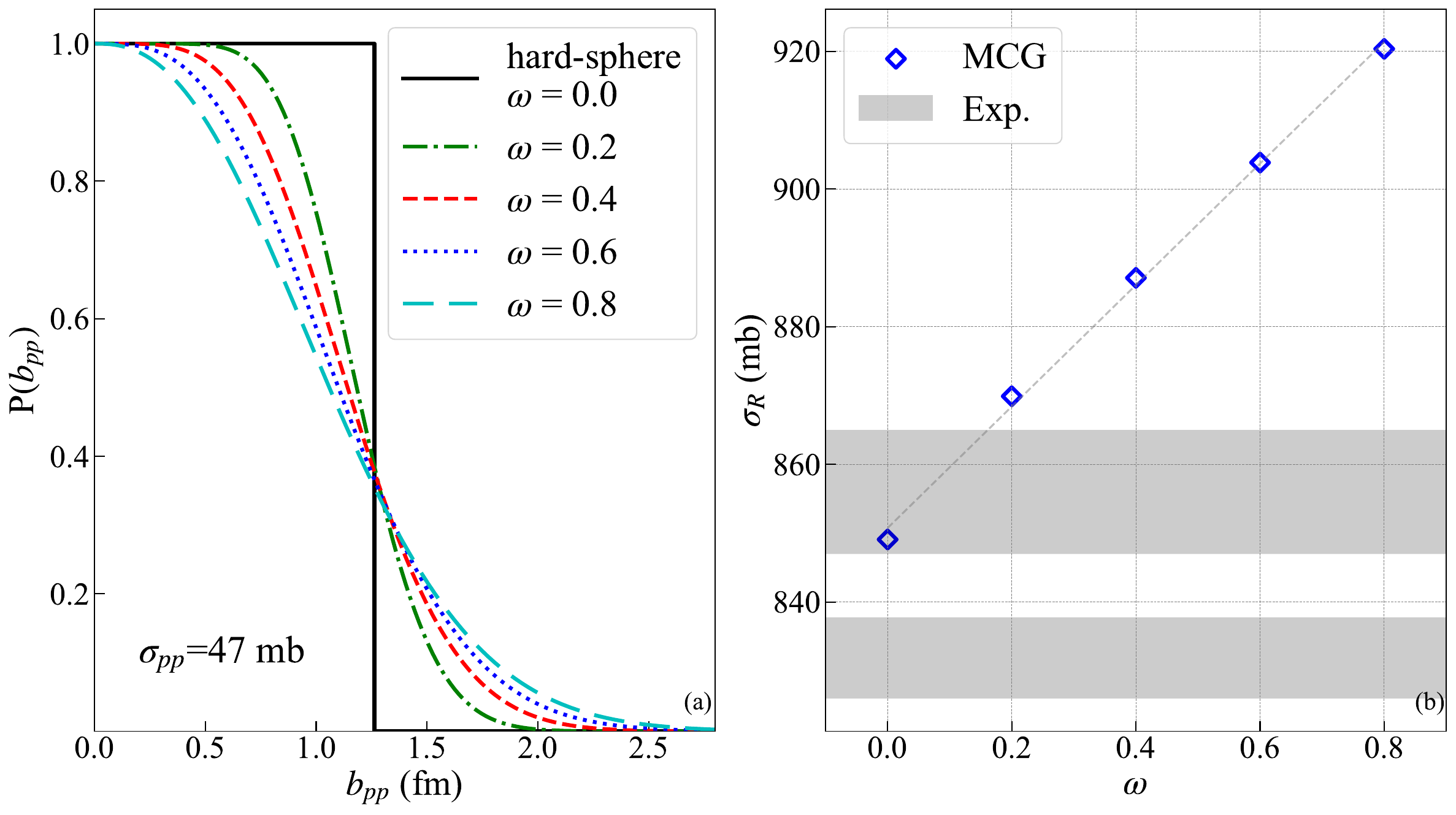}
\caption{
Sensitivity of the MCG calculation to the transverse nucleon--nucleon
collision profile.
(a) Proton--proton collision probability $P(b_{pp})$ for
$\sigma_{pp}=47~\mathrm{mb}$. The solid black curve
denotes the hard-sphere limit ($\omega=0$), while the colored curves
represent smooth profiles with the indicated values of $\omega$.
(b) Calculated reaction cross sections for
$^{12}\mathrm{C}+{}^{12}\mathrm{C}$ at $790~\mathrm{MeV/nucleon}$ as a
function of $\omega$. 
The horizontal gray bands indicate the measured interaction-cross-section
intervals $\sigma_I$ reported at $800$ and
$790~\mathrm{MeV/nucleon}$ in
Refs.~\cite{12Con12CPONNATH2024PLB,12con12COZAWA2001NPA},
respectively.
The dashed line connecting the calculated points is included solely as a
guide to the eye.
}\label{fig:sm_omega_test}
\end{figure}

To assess the sensitivity to the transverse $NN$ collision profile, we vary $\omega$ while keeping the same energy- and isospin-dependent total $NN$ cross sections. Figure~\ref{fig:sm_omega_test} shows the results for $^{12}\mathrm{C}+{}^{12}\mathrm{C}$ at $790~\mathrm{MeV/nucleon}$ as a representative test.

As shown in Fig.~\ref{fig:sm_omega_test}, increasing $\omega$ smooths
the binary $NN$ collision profile. At fixed total $\sigma_{NN}$, the
collision probability is reduced at small transverse separations and
develops a longer tail beyond the hard-sphere cutoff, resulting in a
monotonic increase in $\sigma_R$ for this system.
The two measured cross section data
shown in Fig.~\ref{fig:sm_omega_test}, reported at nearby incident
energies of $800$ and $790~\mathrm{MeV/nucleon}$ in
Refs.~\cite{12Con12CPONNATH2024PLB,12con12COZAWA2001NPA},
respectively, do not overlap within their quoted uncertainties. Among
the profiles examined in this representative test, the hard-sphere
limit ($\omega=0$) overlaps with  the higher data set and lies closest to the
lower one, whereas all tested smooth profiles yield cross
sections above both data sets. We adopt the hard-sphere prescription
as a parameter-free baseline throughout this work; it is fixed entirely
by the energy- and isospin-dependent total $NN$ cross sections and
introduces no additional profile-shape parameter.

\subsection{Nuclear lattice effective field theory} 
\subsubsection{Lattice Hamiltonian}
Nuclear Lattice Effective Field Theory solves the nuclear $A$-body problem in a finite volume by discretizing spacetime on a hypercubic lattice of size $L \times L \times L \times L_t$, with spatial and temporal lattice spacings $a$ and $a_t$, respectively. In this work, we employ the N$^3$LO interaction of Ref.~\cite{Serdarnature2024} with $L=13.2$~fm, $a=1.32~\mathrm{fm}$ and $a_t=0.2~\mathrm{fm}$.

To alleviate the Monte Carlo sign problem, we employ the wave function matching method of Ref.~\cite{Serdarnature2024} and treat the remaining interaction terms perturbatively. The high-fidelity $\chi$EFT Hamiltonian $H$ at N$^3$LO is given by
\begin{equation}
\begin{aligned}
  H
  ={}&
  K
  +V_{\rm OPE}^{\Lambda_\pi=300}
  +V_{\rm C_{\pi}}^{\Lambda_\pi=300}
  +V_{\rm Coulomb}
  +V_{\rm 3N}^{Q^3}
  \\
  &+
  V_{\rm 2N}^{Q^4}
  +W_{\rm 2N}^{Q^4}
  +V_{\rm 2N,WFM}^{Q^4}
  +W_{\rm 2N,WFM}^{Q^4}.
\end{aligned}
\label{eq:H_N3LO}
\end{equation}
Here, $V_{\rm OPE}^{\Lambda_\pi=300}$ and
$V_{\rm C_{\pi}}^{\Lambda_\pi=300}$ denote the one-pion-exchange
(OPE) potential and its counterterm, respectively, both with cutoff
$\Lambda_\pi=300$~MeV. The terms $V_{\rm Coulomb}$,
$V_{\rm 3N}^{Q^3}$, and $V_{\rm 2N}^{Q^4}$ denote the Coulomb
interaction, the three-nucleon (3N) interaction at order $Q^3$, and
the two-nucleon (2N) short-range interaction at order $Q^4$,
respectively. The term $W_{\rm 2N}^{Q^4}$ restores Galilean
invariance, $V_{\rm 2N,WFM}^{Q^4}$ is the wave-function-matching
interaction, and $W^{Q^4}_{\rm 2N,WFM}$ is the corresponding GIR
correction. Details of these terms can be found in
Ref.~\cite{Serdarnature2024}. Note that the two-pion exchange $NN$ potential is absorebd in the short-distance contact terms.

\subsubsection{Pinhole method in perturbation theory}
To formulate the perturbative expansion, we decompose the Hamiltonian
$H$ as
\begin{equation}
  H = H_S + (H - H_S).
\end{equation}
The simple Hamiltonian $H_S$ is used as the nonperturbative part and is
constructed as:
\begin{equation}\label{eq:H_S}
  H_S = K
  + \frac{c_{\rm SU(4)}}{2}\sum_{\bm n}:\!\widetilde{\rho}^{2}(\bm n)\!:
  + V_{\rm OPE}^{\Lambda_\pi=180}.
\end{equation}
Here, $K$ denotes the kinetic energy. The coupling
$c_{\rm SU(4)}$ multiplies the smeared SU(4)-symmetric contact term.
The operators $\widetilde{\rho}$ is the smeared total density, and the colons
denote normal ordering. The difference $H-H_S$ is treated as the
perturbation.

Accordingly, the transfer matrix can be expanded to first order as
\begin{equation}
\begin{aligned}
  M
  ={}& :\!\exp(-a_t H)\!:
  \\
  \simeq{}& :\!\exp(-a_t H_S)\!:
  -a_t:\!\exp(-a_t H_S)(H-H_S)\!:
  \equiv M^{(0)}+M^{(1)}.
\end{aligned}
\label{eq:transfer_matrix_first_order}
\end{equation}

For compactness, we denote a complete $A$-nucleon pinhole
configuration and its normal-ordered density operator by
\begin{equation}
  \mathcal C_A
  =
  \{(\bm n_k,i_k,j_k)\}_{k=1}^{A},
  \qquad
  \rho_A(\mathcal C_A)
  \equiv~
  :\!\prod_{k=1}^{A}\rho_{i_k,j_k}(\bm n_k)\!: .
  \label{eq:pinhole_configuration}
\end{equation}
Here, $\rho_{i,j}(\bm n)$ is the one-body density operator for a
nucleon with spin $i$ and isospin $j$ at lattice site $\bm n$, and
$\sum_{\mathcal C_A}$ denotes the complete sum over all spatial,
spin, and isospin pinhole labels. For a fixed configuration
$\mathcal C_A$, the density operator is inserted at the midpoint of
the Euclidean-time projection amplitude,
\begin{equation}
  \langle\Psi|
  M^{L_t/2}\rho_A(\mathcal C_A)M^{L_t/2}
  |\Psi\rangle ,
  \label{eq:smeq1}
\end{equation}
where $|\Psi\rangle$ is the trial state.

For an observable $O(\mathcal C_A)$ that depends on the pinhole
configuration, the corresponding observable amplitude is
\begin{equation}
  \mathcal M_O
  =
  \sum_{\mathcal C_A}
  \langle\Psi|
  M^{L_t/2}\rho_A(\mathcal C_A)O(\mathcal C_A)M^{L_t/2}
  |\Psi\rangle .
  \label{eq:single_nucleus_observable_amplitude}
\end{equation}
Through first order in $M^{(1)}$, the normalized expectation value is
\begin{equation}
\begin{aligned}
  \langle O\rangle
  ={}  \frac{\mathcal M_O^{(0)}+\mathcal M_O^{(1)}}
       {\mathcal M^{(0)}+\mathcal M^{(1)}}
  ={}  \frac{\mathcal M_O^{(0)}}{\mathcal M^{(0)}}
  +
  \left[
  \frac{\mathcal M_O^{(1)}}{\mathcal M^{(0)}}
  -
  \frac{\mathcal M_O^{(0)}\mathcal M^{(1)}}
       {(\mathcal M^{(0)})^2}
  \right]
  +
  \mathcal O\!\left((\mathcal M^{(1)})^2\right).
\end{aligned}
\label{eq:single_nucleus_first_order}
\end{equation}
The zeroth-order amplitudes are
\begin{align}
  \mathcal M^{(0)}
  &=
  \langle\Psi|(M^{(0)})^{L_t}|\Psi\rangle ,
  \nonumber\\
  \mathcal M_O^{(0)}
  &=
  \sum_{\mathcal C_A}
  \langle\Psi|
  (M^{(0)})^{L_t/2}
  \rho_A(\mathcal C_A)O(\mathcal C_A)
  (M^{(0)})^{L_t/2}
  |\Psi\rangle .
  \label{eq:single_nucleus_zeroth_amplitudes}
\end{align}
The first-order amplitudes are obtained by replacing one occurrence of
$M^{(0)}$ by $M^{(1)}$ and summing over all possible insertion times,
\begin{equation}
\begin{aligned}
  \mathcal M^{(1)}
  ={}&
  \sum_{n_t=0}^{L_t/2-1}
  \langle\Psi|
  (M^{(0)})^{L_t-n_t-1}
  M^{(1)}(M^{(0)})^{n_t}
  |\Psi\rangle
  +\mathrm{c.c.},
  \\
  \mathcal M_O^{(1)}
  ={}&
  \sum_{\mathcal C_A}
  \sum_{n_t=0}^{L_t/2-1}
  \langle\Psi|
  (M^{(0)})^{L_t/2}
  \rho_A(\mathcal C_A)O(\mathcal C_A)\times
  (M^{(0)})^{L_t/2-n_t-1}
  M^{(1)}(M^{(0)})^{n_t}
  |\Psi\rangle
  +\mathrm{c.c.}.
\end{aligned}
\label{eq:single_nucleus_first_amplitudes}
\end{equation}
All four amplitudes can be calculated using the auxiliary-field
formalism~\cite{LU2019PLB}.

\subsubsection{Perturbation calculation involving two independent nuclei}
The adiabatic approximation is applicable to nuclear collisions at
intermediate to high energies~\cite{Suzuki2003}. We therefore write
the projectile--target trial state as the direct product
\begin{equation}
  |\widetilde\Psi\rangle
  =
  |\Psi^{\rm proj}\rangle\otimes|\Psi^{\rm targ}\rangle .
  \label{eq:projectile_target_trial_state}
\end{equation}
The complete projectile and target pinhole configurations are denoted
by
\begin{equation}
\begin{aligned}
  \mathcal C_P
  &=
  \{(\bm n_{P,k},i_{P,k},j_{P,k})\}_{k=1}^{A_P},
  \\
  \mathcal C_T
  &=
  \{(\bm n_{T,k},i_{T,k},j_{T,k})\}_{k=1}^{A_T}.
\end{aligned}
\label{eq:projectile_target_pinhole_configurations}
\end{equation}
The operators $\rho_{A_P}^{\rm proj}(\mathcal C_P)$ and
$\rho_{A_T}^{\rm targ}(\mathcal C_T)$ below are the projectile and
target versions of the $A$-body density operator in
Eq.~\eqref{eq:pinhole_configuration}.
For a joint observable
$\widetilde O(\mathcal C_P,\mathcal C_T)$, the direct observable
amplitude is
\begin{equation}
\begin{aligned}
  \widetilde{\mathcal M}_O
  ={}  \sum_{\mathcal C_P}\sum_{\mathcal C_T}
  \langle\widetilde\Psi|
  M_{\rm proj}^{L_t/2}M_{\rm targ}^{L_t/2}\times
  \rho_{A_P}^{\rm proj}(\mathcal C_P)
  \rho_{A_T}^{\rm targ}(\mathcal C_T)
  \widetilde O(\mathcal C_P,\mathcal C_T)\times
  M_{\rm proj}^{L_t/2}M_{\rm targ}^{L_t/2}
  |\widetilde\Psi\rangle .
\end{aligned}
\label{nonpert}
\end{equation}
The corresponding normalized expectation value is
\begin{equation}
  \langle\widetilde O\rangle_{\rm direct}
  =
  \frac{\widetilde{\mathcal M}_O}
       {\mathcal M_{\rm proj}\mathcal M_{\rm targ}} ,
  \label{eq:two_nucleus_direct}
\end{equation}
where
\begin{equation}
  \mathcal M_\alpha
  =
  \langle\Psi^\alpha|M_\alpha^{L_t}|\Psi^\alpha\rangle ,
  \qquad
  \alpha\in\{{\rm proj},{\rm targ}\}.
  \label{eq:projectile_target_normalizations}
\end{equation}
The normalization amplitude factorizes because the projectile and
target are propagated independently. The observable amplitude does
not generally factorize, because
$\widetilde O(\mathcal C_P,\mathcal C_T)$ depends simultaneously on
the two configurations, it is therefore retained as a joint
projectile--target amplitude.

We denote by $\widetilde{\mathcal M}_{O}^{(0)}$ the joint observable
amplitude with both nuclei propagated at zeroth order. The amplitudes
$\widetilde{\mathcal M}_{O,{\rm proj}}^{(1)}$ and
$\widetilde{\mathcal M}_{O,{\rm targ}}^{(1)}$ are obtained by
replacing one projectile or target transfer matrix $M^{(0)}$ by
$M^{(1)}$, respectively, summing over all possible insertion times,
and keeping the other nucleus at zeroth order.
The normalization amplitudes $\mathcal M_\alpha^{(0)}$ and
$\mathcal M_\alpha^{(1)}$ are the corresponding single-nucleus
amplitudes defined as in
Eqs.~\eqref{eq:single_nucleus_zeroth_amplitudes} and
\eqref{eq:single_nucleus_first_amplitudes}.

Keeping all terms through first order in the total perturbation gives
\begin{equation}
\begin{aligned}
  \langle\widetilde O\rangle_{\rm 1st}
  ={}\frac{\widetilde{\mathcal M}_{O}^{(0)}}
       {\mathcal M_{\rm proj}^{(0)}
        \mathcal M_{\rm targ}^{(0)}}+
  \frac{
  \widetilde{\mathcal M}_{O,{\rm proj}}^{(1)}
  +
  \widetilde{\mathcal M}_{O,{\rm targ}}^{(1)}
  }{
  \mathcal M_{\rm proj}^{(0)}
  \mathcal M_{\rm targ}^{(0)}
  }-  \frac{\widetilde{\mathcal M}_{O}^{(0)}}
       {\mathcal M_{\rm proj}^{(0)}
        \mathcal M_{\rm targ}^{(0)}}
  \left(
  \frac{\mathcal M_{\rm proj}^{(1)}}
       {\mathcal M_{\rm proj}^{(0)}}
  +
  \frac{\mathcal M_{\rm targ}^{(1)}}
       {\mathcal M_{\rm targ}^{(0)}}
  \right)
  +
  \mathcal O\!\left((\mathcal M^{(1)})^2\right).
\end{aligned}
\label{eq:two_nucleus_first_order}
\end{equation}
Here, $\mathcal O[(\mathcal M^{(1)})^2]$ collectively denotes all
second- and higher-order contributions in the perturbative insertion.
In particular,
$\mathcal M_{\rm proj}^{(1)}\mathcal M_{\rm targ}^{(1)}$ contains one
first-order insertion in each nucleus and is therefore second order
in the total perturbation.

For fixed projectile and target orientations and a fixed
impact-parameter vector $\mathbf b$, we choose
\begin{equation}
  \widetilde O(\mathcal C_P,\mathcal C_T;\mathbf b)
  =
  I_R(\mathbf b;\mathcal C_P,\mathcal C_T),
  \label{eq:reaction_indicator_observable}
\end{equation}
where $I_R$ is the reaction indicator of
Eq.~\eqref{eq:sm_mcg_event_indicator} for the event specified by the
two pinhole configurations and their orientations. The external orientation and impact-parameter azimuthal averages are then performed as in Eq.~\eqref{eq:sm_mcg_probability_average}. For all production calculations reported in the main text, the reaction
probability $P_R^{\rm MCG}(b)$ is evaluated using the strict first-order
expectation value $\langle\widetilde O\rangle_{\rm 1st}$ in
Eq.~\eqref{eq:two_nucleus_first_order}, and the reaction cross section
is subsequently obtained from Eq.~\eqref{eq:sm_mcg_sigma}. The direct
normalized expression in Eq.~\eqref{eq:two_nucleus_direct} is used only
for the verification test described below.

As a verification test, we consider the simplified SU(4)
Hamiltonian of Ref.~\cite{LU2019PLB},
\begin{equation}
  H_{\rm SU(4)}
  =
  H_{\rm free}
  +
  \frac{C_2}{2!}\sum_{\bm n}\widetilde\rho(\bm n)^2
  +
  \frac{C_3}{3!}\sum_{\bm n}\widetilde\rho(\bm n)^3 .
  \label{eq:HSU4}
\end{equation}
Here, $H_{\rm free}$ is the free Hamiltonian, $C_2$ and $C_3$ are
the two- and three-body couplings, and $\widetilde\rho$ is the
smeared SU(4)-invariant density. For the direct calculation, we set
\begin{equation}
  C_2^{\rm dir}
  =
  -3.75\times10^{-7}~\mathrm{MeV}^{-2}
\end{equation}
and use Eqs.~\eqref{nonpert} and
\eqref{eq:two_nucleus_direct}. For the strict first-order
calculation, the unperturbed coupling is
\begin{equation}
  C_2^{(0)}
  =
  -3.41\times10^{-7}~\mathrm{MeV}^{-2},
\end{equation}
and the perturbation is
\begin{equation}
  \Delta H
  =
  \frac{\Delta C_2}{2!}
  \sum_{\bm n}\widetilde\rho(\bm n)^2,
  \qquad
  \Delta C_2
  =
  -0.34\times10^{-7}~\mathrm{MeV}^{-2},
  \label{eq:su4_benchmark_perturbation}
\end{equation}
so that $C_2^{(0)}+\Delta C_2=C_2^{\rm dir}$. The value of $C_3$,
the smearing parameters, the incident energy, and all MCG reaction
settings are held fixed between the two calculations. For
$^{12}\mathrm C+{}^{12}\mathrm C$, the first-order result from
Eq.~\eqref{eq:two_nucleus_first_order} is
$\sigma_R^{\rm 1st}=685~\mathrm{mb}$, whereas the direct
calculation gives $\sigma_R^{\rm dir}=699~\mathrm{mb}$. The two
results differ by approximately $2\%$ in this benchmark.

\bibliography{References}
\bibliographystyle{apsrev}